\documentclass[aps,twocolumn,showpacs,preprintnumbers,amsmath,amssymb,superscriptaddress,floatfix,nofootinbib]{revtex4}

\usepackage{graphicx}
\usepackage{epsfig}
\usepackage{epstopdf}
\usepackage{hyperref}
\usepackage{amsmath}
\usepackage{amsfonts}
\usepackage{amssymb}

\usepackage{graphicx,color,dcolumn,booktabs}
\usepackage{txfonts}
\usepackage{amssymb}
\usepackage{indentfirst}
\usepackage{subfig}
\usepackage{float}

\allowdisplaybreaks
\begin{document}

\title{$D_{s0}^{*}(2317)^+$ in the decay of $B_c$ into $J/\psi DK$}


\author{Zhi-Feng Sun}
\affiliation{Departamento de
F\'{\i}sica Te\'orica and IFIC, Centro Mixto Universidad de
Valencia-CSIC Institutos de Investigaci\'on de Paterna, Aptdo.
22085, 46071 Valencia, Spain}

\author{M. Bayar}
\affiliation{Department of Physics, Kocaeli University, 41380 Izmit, Turkey}

\author{P.~ Fernandez-Soler}
\affiliation{Departamento de
F\'{\i}sica Te\'orica and IFIC, Centro Mixto Universidad de
Valencia-CSIC Institutos de Investigaci\'on de Paterna, Aptdo.
22085, 46071 Valencia, Spain}

\author{E.~Oset}
\affiliation{Departamento de
F\'{\i}sica Te\'orica and IFIC, Centro Mixto Universidad de
Valencia-CSIC Institutos de Investigaci\'on de Paterna, Aptdo.
22085, 46071 Valencia, Spain}

\date{\today}

\begin{abstract}
In this paper we study the relationship between the $D_{s0}^{*}(2317)^+$ resonance and the decay of the $B_c$ meson into $J/\psi DK$. In this process, the $B_c$ meson decays first into $J/\psi$ and the quark pair $c\bar{s}$, and then the quark pair hadronizes into $DK$ or $D_s\eta$ components, which undergo final state interaction. This final state interaction, generating the $D_{s0}^{*}(2317)^+$ resonance, is described by the chiral unitary approach. With the parameters which allow us to match the pole position of the $D_{s0}^{*}(2317)^+$, we obtain the $DK$ invariant mass distribution of the decay $B_c\to J/\psi DK$, and also the rate for $B_c\to J/\psi D_{s0}^{*}(2317)$. The ratio of these two magnitude is then predicted.
\end{abstract}

\maketitle

\section{Introduction}
In 1998, the discovery of the $B_c$ was reported by the CDF collaboration in the $B_c\to J/\psi l^{\pm}\bar{v}_l$ process at Fermilab \cite{Abe:1998wi}. About 10 years later, the CDF collaboration confirmed the existence of $B_c$ via the decay mode $B_c^{\pm}\to J/\psi \pi^{\pm}$ at CDF II with a significance of more than $8\sigma$ \cite{Aaltonen:2007gv}. Additionally, the D0 collaboration observed the $B_c$ states in the $B_c^{\pm}\to J/\psi \pi^{\pm}$ process with a significance of more than $5\sigma$ \cite{Abazov:2008kv}. In the last two years, the $B_c$ state has also been observed by the LHCb collaboration with the decay modes $B_c^{\pm}\to J/\psi \pi^{\pm}$ and $B_c^{\pm}\to J/\psi D_s^{\pm}$ at the LHC center-of-mass energy 7 TeV of proton-proton collisions {\cite{Aaij:2012dd, Aaij:2013gia}}. The average value for the mass of the $B_c$ state listed in the Particle Data Group (PDG) is $M_{B_c}=6.2756 \pm 0.0011$ GeV \cite{Agashe:2014kda}.

Another state related to our work is the $D_{s0}^{*}(2317)^+$ resonance, of which we will give a brief review next.

The scalar resonance $D_{s0}^{*}(2317)^+$ was first observed by BABAR Collaboration as a narrow peak in the inclusive $e^+e^-\to D_s^+\pi^0X$ annihilation process \cite{Aubert:2003fg, Aubert:2003pe}. Later, this observation was confirmed by CLEO, BELLE and FOCUS Collaborations \cite{Besson:2003cp, Krokovny:2003zq, Vaandering:2004ix}. The average mass of $D_{s0}^{*}(2317)^+$ listed in the PDG is $2318.0\pm 1.0$ MeV \cite{Agashe:2014kda}. 

Before the BABAR experiment, the potential model \cite{Godfrey:1985xj, Godfrey:1986wj, Gupta:1994mw, Zeng:1994vj, Ebert:1997nk, DiPierro:2001uu, Kalashnikova:2001ig, Merten:2001er, Lucha:2003gs}
and lattice QCD \cite{Boyle:1997rk, Bali:2003jv, Dougall:2003hv} studied the P-wave charmed strange meson and predicted a meson mass larger than the experimental value, and a width of the $D_{s0}^{*}(2317)^+\to DK$ decay very large. After the BABAR experiment, many theoretical groups performed research on the $D_{s0}^{*}(2317)^+$ state.
Since the mass of the $D_{s0}^{*}(2317)^+$ is close to the threshold of the $DK$ system, being the difference of about 50 MeV, the molecular state interpretation was proposed \cite{Barnes:2003dj, Szczepaniak:2003vy, Kolomeitsev:2003ac, Guo:2006fu, Gamermann:2006nm, Guo:2009ct, Cleven:2010aw, Cleven:2014oka, Faessler:2007cu, Faessler:2007gv}.
In Refs. \cite{vanBeveren:2003kd, Cheng:2003kg, Terasaki:2003qa, Maiani:2004vq, Bracco:2005kt, Dmitrasinovic:2005gc, Browder:2003fk}, the $D_{s0}^{*}(2317)^+$ state was studied in the frame of $KD$ mixing with $c\bar{s}$ state, four-quark state, and the mixture of two-meson and four-quark state.
In Refs. \cite{Mohler:2013rwa, Lang:2014yfa}, introducing $KD$ meson operators and using the effective range formula, the authors obtained a bound state about $40$ MeV below the $KD$ threshold, which was reanalysed in Ref. \cite{Torres:2014vna}. In Ref. \cite{Liu:2012zya}, lattice QCD results for the $DK$ scattering length were extrapolated to physical pion masses by means of unitarized chiral perturbation theory, and by means of the Weinberg compositeness condition \cite{Weinberg:1965zz, Baru:2003qq} the amount of $KD$ content in the $D_{s0}^{*}(2317)^+$ was determined, resulting in a sizable fraction of the order of $70 \%$ within errors. \footnote{However, as discussed in detail in \cite{Torres:2014vna}, the Weinberg compositeness condition was used in an extreme case in \cite{Liu:2012zya}, thus weakening the conclusions about this fraction. A more accurate determination of that fraction, of the order of $70\%$ is done in \cite{Torres:2014vna}.}

In the present work, we shall give the $DK$ invariant mass distribution in the decay $B_c\to J/\psi DK$, from which information on the internal structure of the $D_{s0}^{*}(2317)^+$ state will be obtained. Besides the weak decay of the $B_c$ meson and hadronization of the quark-antiquark pair to two mesons, the final state interaction is involved. In order to describe the final state interaction, by which the $D_{s0}^{*}(2317)^+$ state is generated, we use the chiral unitary approach which makes use of the on-shell version of the factorized Bethe-Salpeter equation which has successfully explained the existence of some resonances (see \cite{Kaiser:1995eg, Oset:1997it, Oller:2000fj, Hyodo:2002pk, Jido:2003cb, GarciaRecio:2005hy, Molina:2008jw, Geng:2008gx, Molina:2009eb, Gonzalez:2008pv, Sarkar:2009kx, Oset:2009vf}). 

The paper is organized as follows. In Section II we present the formalism to study the decay of $B_c\to J/\psi DK$ and $B_c\to J/\psi D_{s0}^*(2317)$. The numerical results of the $DK$ invariant mass distribution are given in Section III. Finally, we present a brief conclusion.

\section{Formalism}
\begin{figure}[t]
\begin{tabular}{c}
\includegraphics[scale=0.35]{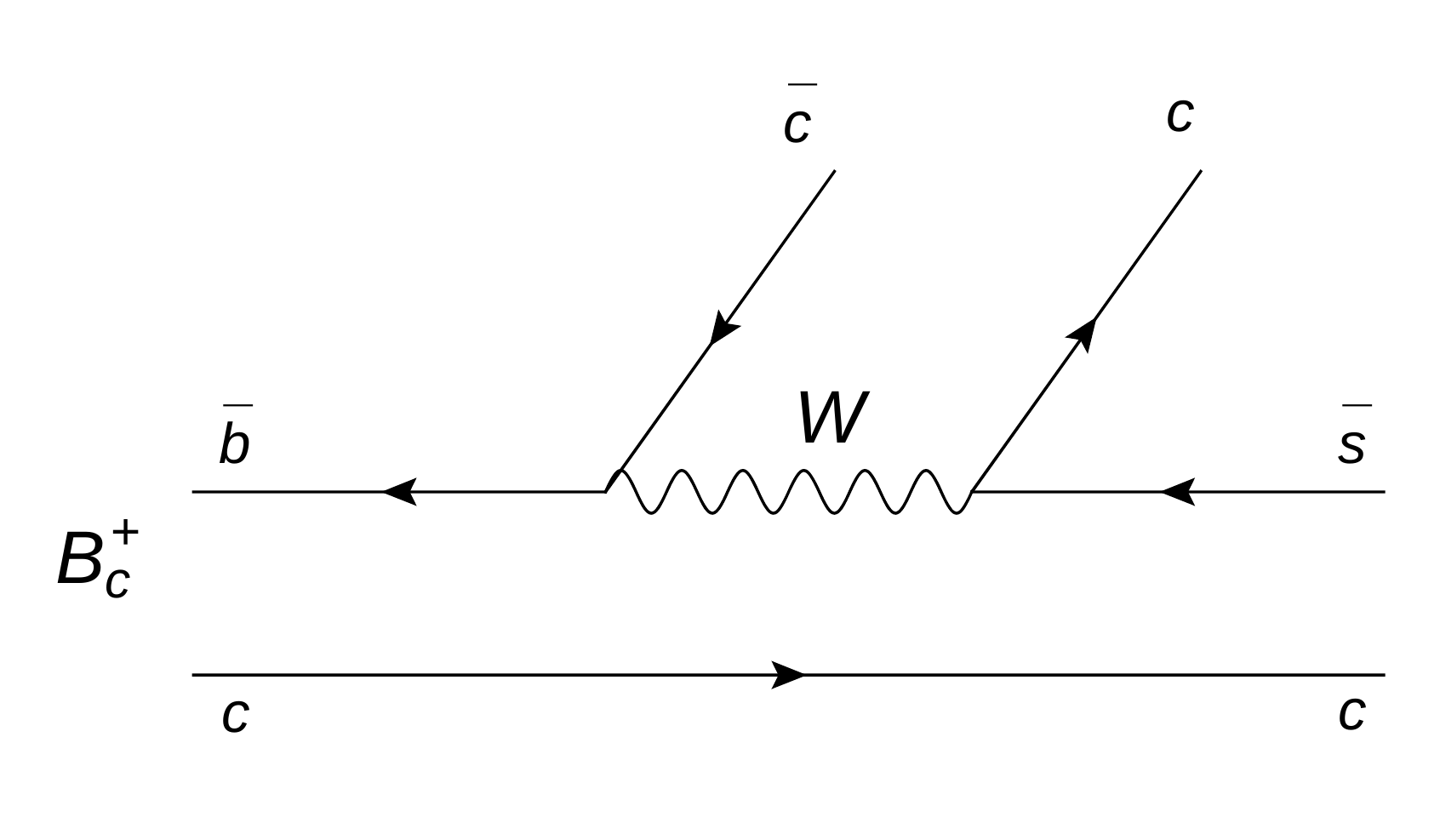}
\end{tabular}
\caption{Diagram for the decay of $B_c^+$ into $J/\psi$ and the quark pair $c\bar{s}$ with internal emission.\label{fig:Bc_to_Jpsi_csbar}}
\end{figure}

\begin{figure}[t]
\begin{tabular}{c}
\includegraphics[scale=0.35]{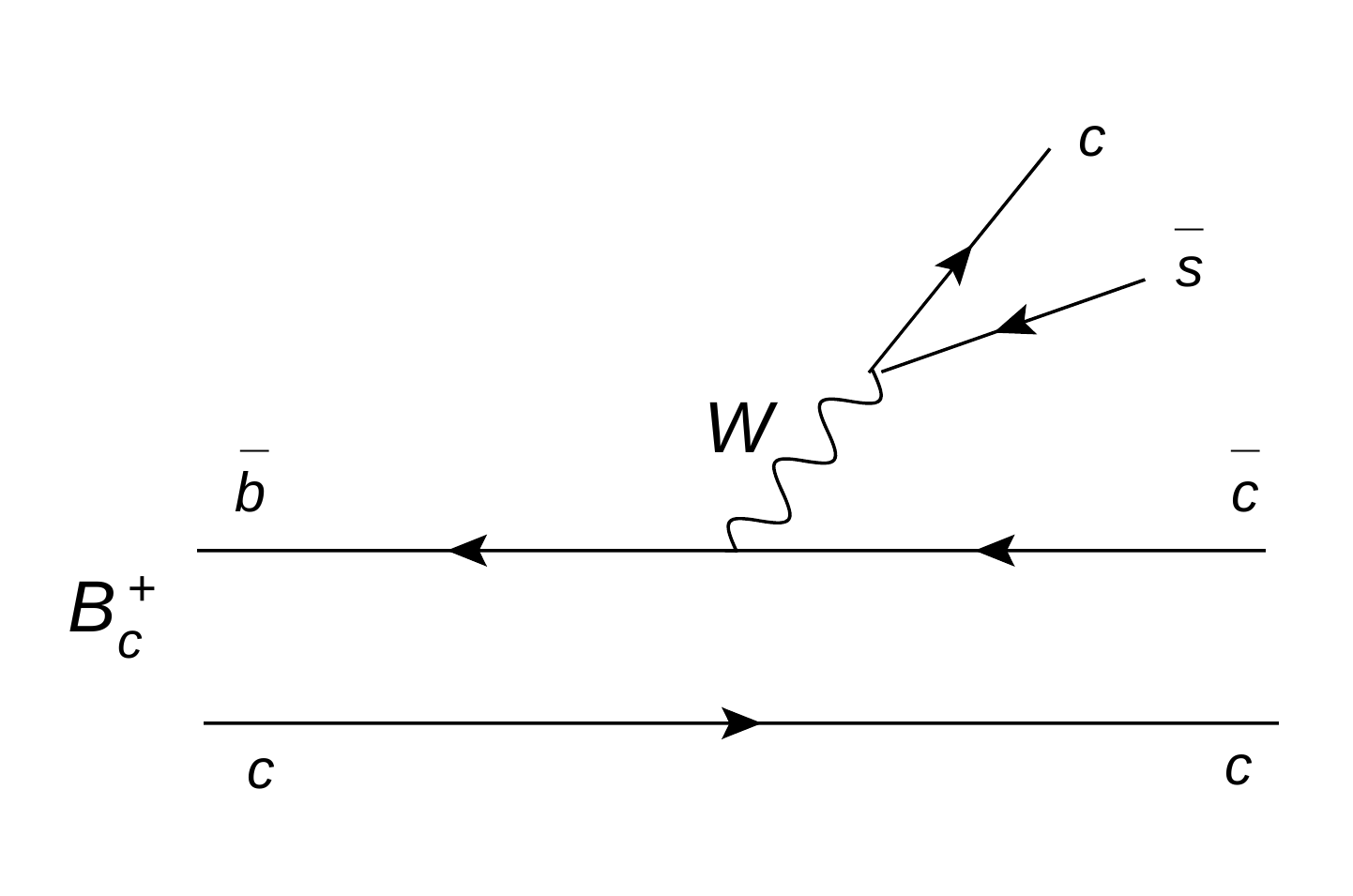}
\end{tabular}
\caption{Diagram for the decay of $B_c^+$ into $J/\psi$ and the quark pair $c\bar{s}$ with external emission.\label{fig:Bccsbarccbar}}
\end{figure}

In this paper, we will discuss the decay mechanism of the $B_c$ meson into $J/\psi DK$ and also into $J/\psi D_{s0}^*(2317)$. In Refs. \cite{Liang:2014tia, Bayar:2014qha, Xie:2014gla, Liang:2014ama, Albaladejo:2015kea}, the weak decay mechanisms of the $B$ and $B_s$ mesons were studied. 
We can take many elements from those works, but there are also some important differences. The works of \cite{Liang:2014tia, Bayar:2014qha} relied upon the topological diagrams of Fig. \ref{fig:Bc_to_Jpsi_csbar}, which we have adapted to the present problem. Essentially a $d$ quark from the $B^0$ meson is replaced now by a $c$ quark in the $B_c^+$ case here. This diagram is addressed as internal emission in the nomenclature of Ref. \cite{Chau:1982da, Cheng:2010vk}. However we can also have a mechanism of external emission as depicted in Fig. \ref{fig:Bccsbarccbar}.

\begin{figure}[b]
\begin{tabular}{c}
\includegraphics[scale=0.35]{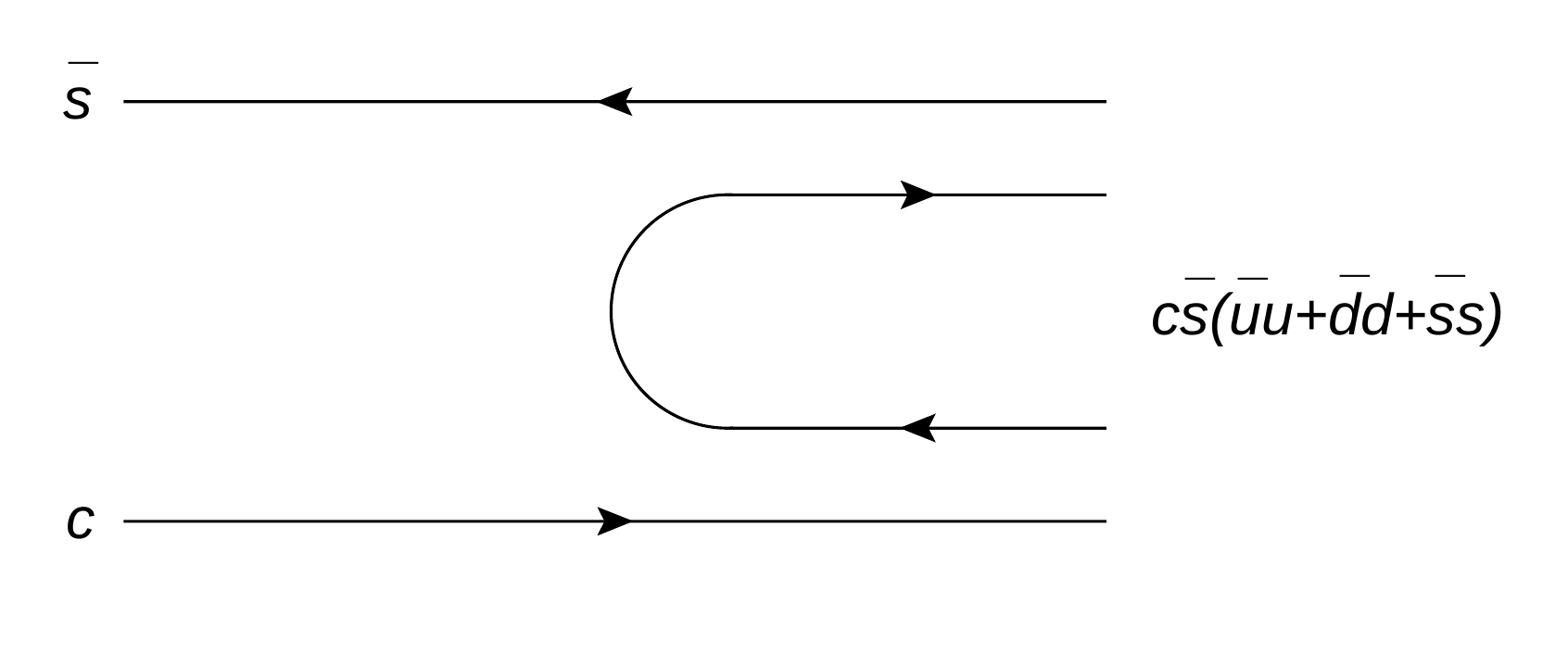}
\end{tabular}
\caption{The hadronization of $c\bar{s}\to c\bar{s}(u\bar{u}+d\bar{d}+s\bar{s})$.\label{fig:csbar_to_DK}}
\end{figure}

\begin{figure*}[htb]
\begin{tabular}{c}
\includegraphics[scale=0.8]{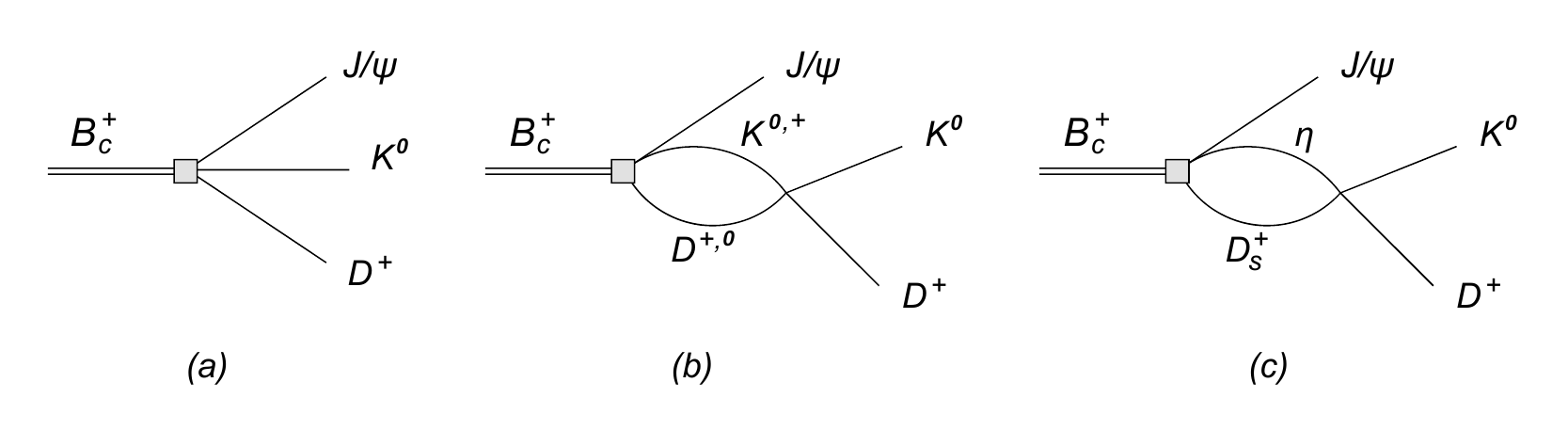}
\end{tabular}
\caption{The diagrams of the decay $B_c^+\to J/\psi D^+K^0$ at hadronic level.\label{fig:Bc_to_Jpsi_DK}}
\end{figure*}

The diagram of Fig. \ref{fig:Bccsbarccbar} is colored favoured and dominates the transition, but in both cases we have a primary $J/\psi c\bar{s}$ production assuming a $c\bar{c}$ pair combining into $J/\psi$. This is all that we need in the present case, since the matrix element for this transition will be factorized and assumed to be constant in the small range of the $KD$ invariant mass that we need in our problem. The smoothness of the weak plus hadronization form factors is supported by calculations \cite{Kang:2013jaa} and phenomenology (see a detailed discussion in \cite{Sekihara:2015iha}). The next step consists of the hadronization of the $c\bar{s}$ pair into two mesons. This is depicted in Fig. \ref{fig:csbar_to_DK} and is implemented introducing an extra $\bar{q}q$ pair with the quantum numbers of the vacuum, $\bar{u}u+\bar{d}d+\bar{s}s$. In a third step, the two mesons produced in the second process may interact with themselves in coupled channels, which is shown in Fig. \ref{fig:Bc_to_Jpsi_DK}. 


It should be noted that apart from the necessary $bc$ transition in the weak process, the other weak vertex is $V_{cs}$ in both mechanisms. As mentioned before, we will include the matrix elements for the weak plus hadronization processes into a constant factor that we call $V_p$.
The $c\bar{s}$ then recombine with the $q\bar{q}$ created from the vacuum producing two mesons. In order to calculate it, we first consider the $q\bar{q}$ matrix $M$
\begin{eqnarray}
M=\left(
\begin{array}{cccc}
u\bar{u}&u\bar{d}&u\bar{s}&u\bar{c}\\
d\bar{u}&d\bar{d}&d\bar{s}&d\bar{c}\\
s\bar{u}&s\bar{d}&s\bar{s}&s\bar{c}\\
c\bar{u}&c\bar{d}&c\bar{s}&c\bar{c}\\
\end{array}\right)=\left(
\begin{array}{cccc}
u\\
d\\
s\\
c\\
\end{array}\right)\left(
\begin{array}{cccc}
\bar{u}&\bar{d}&\bar{s}&\bar{c}\\
\end{array}\right),
\end{eqnarray}
which has the property,
\begin{eqnarray}
M\cdot M&=&\left(
\begin{array}{cccc}
u\\
d\\
s\\
c\\
\end{array}\right)\left(
\begin{array}{cccc}
\bar{u}&\bar{d}&\bar{s}&\bar{c}\\
\end{array}\right)\left(
\begin{array}{cccc}
u\\
d\\
s\\
c\\
\end{array}\right)\left(
\begin{array}{cccc}
\bar{u}&\bar{d}&\bar{s}&\bar{c}\\
\end{array}\right)\nonumber\\
&=&\left(
\begin{array}{cccc}
u\\
d\\
s\\
c\\
\end{array}\right)\left(
\begin{array}{cccc}
\bar{u}&\bar{d}&\bar{s}&\bar{c}\\
\end{array}\right)(\bar{u}u+\bar{d}d+\bar{s}s+\bar{c}c)\nonumber\\
&=&M\times (\bar{u}u+\bar{d}d+\bar{s}s+\bar{c}c).
\end{eqnarray}

On the hadron level, the $M$ matrix corresponds to the $\phi$ matrix, which has the form,
\begin{eqnarray}
\phi=\left(
\begin{array}{cccc}
\frac{\eta}{\sqrt{3}}+\frac{\pi^0}{\sqrt{2}}+\frac{\eta^\prime}{\sqrt{6}}&\pi^+ & K^+ &\bar{D}^0\\
\pi^- & \frac{\eta}{\sqrt{3}}-\frac{\pi^0}{\sqrt{2}}+\frac{\eta^\prime}{\sqrt{6}} &K^0 &{D}^-\\
K^-&\bar{K}^0 &\frac{\sqrt{2}\eta^\prime}{\sqrt{3}}-\frac{\eta}{\sqrt{3}}&D_s^-\\
D^0 &D^+ &D_s^+&\eta_c\\
\end{array}\right),
\end{eqnarray}
where the standard $\eta-\eta^\prime$ mixing is used \cite{Bramon:1992kr}. 

Then, we get
\begin{eqnarray}
(\bar{u}u+\bar{d}d+\bar{s}s+\bar{c}c)(c\bar{s})\to 
(\phi\phi)_{43}&=&\eta_cD_s^++D^0K^++D^+K^0\nonumber\\
&&-\frac{1}{\sqrt{3}}\eta D_s^++\sqrt{\frac{2}{3}}D_s^+\eta^\prime.\nonumber\\
\end{eqnarray}
In this paper, we neglect the contribution of $\eta^\prime$ and $\eta_c$ because of their large mass compared with the $K$ and $\eta$ masses.

\subsection{Rescattering}
As it is shown in Fig. \ref{fig:Bc_to_Jpsi_DK} (b) and (c), the two mesons produced from the $(c\bar{s})$ (see Fig. \ref{fig:csbar_to_DK}) may interact with themselves and the coupled channels. The amplitude of the $B^+_c \to J/\psi D^+K^0$ decay is 
\begin{eqnarray}
{t}(B^+_c \to J/\psi D^+K^0)=V_p\left(h_1+\sum_i h_iG_{i}t_{i1}\right).\label{eq:am_3body}
\end{eqnarray}
Here $i=1$, $2$, $3$ which label the channels $D^+K^0$, $D^0K^+$ and  $D_s^+\eta$ respectively.
In Eq. \eqref{eq:am_3body}, $h_1=h_2=1$ and $h_3=-\frac{1}{\sqrt{3}}$. 
$G_i$ is the loop function of two meson propagators
\begin{eqnarray}
G_i=i\int \frac{d^4q}{(2\pi)^4}\frac{1}{(P-q)^2-m_{i}^2+i\epsilon}\frac{1}{q^2-M_{i}^2+i\epsilon},
\end{eqnarray}
where $P$ is the total four momentum of the system, and $m_{i}$ and $M_{i}$ are the masses of the mesons in the $i$-channel. 

The loop function is calculated using dimensional regularization, and the function is renormalized by means of a subtraction constant $\alpha(\mu)$. The expression of the calculated loop function is 
\begin{eqnarray}
G_i&=& \frac{1}{16\pi^2}\left(\alpha(\mu)+\log \frac{m_{i}^2}{\mu^2}+\frac{M_{i}^2-m_{i}^2+s}{2s}\log \frac{M_{i}^2}{m_{i}^2} \right.\nonumber\\
&&+\frac{p}{\sqrt{s}}\left(\log \frac{s-M_{i}^2+m_{i}^2+2p\sqrt{s}}{-s+M_{i}^2-m_{i}^2+2p\sqrt{s}} \right.\nonumber\\
&&+\left.\left.\log \frac{s+M_{i}^2-m_{i}^2+2p\sqrt{s}}{-s-M_{i}^2+m_{i}^2+2p\sqrt{s}}\right)\right).\label{eq:loopfunction}
\end{eqnarray}
Here $s$ is the invariant mass squared of the particles appearing in the loop, and $p$ is the corresponding three momentum in the center-of-mass frame. In Eq. (\ref{eq:am_3body}) $t_{ij}$ is the scattering matrix element for the transition channel $i\to j$. According to the on-shell version of the factorized Bethe-Salpeter equation \cite{Oller:2000fj, Oller:1997ti}, $t_{ij}$ is,
\begin{eqnarray}
t=[1-VG]^{-1}V,\label{eq:BSequation}
\end{eqnarray}
where $V$ is the potential which we take from \cite{Gamermann:2006nm}. $V$ is expressed as
\begin{eqnarray}
V_{11}&=&-\frac{1}{6f^2}\left[\frac{3}{2}s-(\gamma+\frac{1}{2})\frac{(m_D^2-m_K^2)^2}{s}\right],\nonumber\\
V_{13}&=&-\frac{1}{6\sqrt{6}f^2}\left[-\frac{3}{2}(\gamma+3)s+\frac{1}{2}(\gamma+3)(m_K^2+m_\eta^2+m_D^2\right.\nonumber\\ 
&&+m_{D_s}^2)-\frac{3}{2}(\gamma+1)\frac{(m_{D_s}^2-m_\eta^2)(m_K^2-m_D^2)}{s}-m_D^2\nonumber\\ 
&&\left.-3m_K^2+2m_\pi^2\right],\nonumber\\ 
V_{33}&=&-\frac{1}{9f^2}\left[-\frac{3}{2}\gamma s+\gamma (m_\eta^2+m_{D_s}^2)-\frac{3}{2}\gamma \frac{(m_{D_s}^2-m_\eta^2)^2}{s} \right.\nonumber\\ 
&& \left.+2(m_D^2+3m_K^2-2m_\pi^2)\right],\nonumber\\
V_{12}&=&V_{11}=V_{22},\nonumber\\ 
V_{23}&=& V_{13},\nonumber\\
V_{ji}&=&V_{ij},\label{eq:potential}
\end{eqnarray}
where $i,j=1,2,3$ and $i\neq j$. And in the above equations, $\gamma=\left(\frac{m_L}{m_H}\right)^2$ with $m_L=800$ MeV and $m_H=2050$ MeV (see Ref. \cite{Gamermann:2006nm}), $f=93$ MeV is the pion decay constant. Note that in Eq. (\ref{eq:potential}) we have projected the potentials into S-wave.

Since the process depicted in Fig. \ref{fig:Bc_to_Jpsi_csbar} is a $0^-\to 1^-0^+$ transition, the angular momentum between the $J/\psi$ and the quark pair $(c\bar{s})$ is $L=1$ due to the total angular momentum conservation. So $V_p$ should have the form of 
\begin{eqnarray}
V_p=\sqrt{3}A p_{J/\psi}\cos \theta.\label{eq:Vp} 
\end{eqnarray}
Thus, we can get the expression of $d\Gamma/dM_{inv}$
\begin{eqnarray}\label{eq::gamma:m}
\frac{d\Gamma}{dM_{inv}}=\frac{A^2}{(2\pi)^3}\frac{1}{4m_{B_c}^2}p_{J/\psi}^3\tilde{p}_{DK}\bar{\sum}\sum\vert \tilde{t}_{B^+_c\to J/\psi D^+K^0}\vert^2, \label{eq:dG/dM}
\end{eqnarray}
where $M_{inv}$ is the invariant mass of the $D^+K^0$ system, and $\tilde{t}_{B^+_c\to J/\psi D^+K^0}$ is $t_{B^+_c\to J/\psi D^+K^0}/{V_p}$. In Eq. (\ref{eq::gamma:m}), the factor $\frac{1}{3}$ which comes from the integral of $\cos \theta^2$ {cancels the $\sqrt{3}^2$ in the definition of $V_p$ of Eq. \eqref{eq:Vp}}. The value of $A$ is chosen to normalize the invariant mass distribution and it will cancel in the ratios that we shall construct. In Eq. (\ref{eq:dG/dM}) $p_{J/\psi}$ is the momentum of the $J/\psi$ in the global CM frame and $\tilde{p}_{DK}$ is the kaon momentum in the $D^+K^0$ rest frame.

It is interesting to see microscopically how a p-wave for $J/\psi$ and the $KD$ meson pair in relative s-wave can be produced in this process. For this we just recall that the $W$ exchange involves the interaction term  $\bar{q}_1\gamma_\mu(1-\gamma_5)q_2\bar{q}_3\gamma^\mu(1-\gamma_5)q_4$ \cite{Buchalla:1995vs, ElBennich:2009da}. By looking at Fig. \ref{fig:Bccsbarccbar}, the required combination is $\gamma^0\gamma_5$ for the $bcW$ vertex, which gives a contribution of the type $\vec{\sigma}\cdot\vec{q}$, which can flip spin to produce $J/\psi$ and also provides the needed p-wave. The vertex for $csW$ should be $\gamma^0$, since sandwiched between the quark and antiquark provides again a vertex of the $\frac{\vec{\sigma}\cdot \vec{p}_1}{2m_1}+\frac{\vec{\sigma}\cdot \vec{p}_2}{2m_2}$ type ($\vec{p}_1$, $\vec{p}_2$ the momenta of the quark, antiquark). Then, upon hadronization of the extra $q\bar{q}$ we can get two mesons in relative s-wave. To see this, recall that the $q\bar{q}$ pair created with the vacuum quantum numbers is produced with spin $S=1$ and $L=1$ ($^3P_0$ configuration \cite{Micu,Close}). The combination of these two p-wave vertices can then give rise to an s-wave of the pair of mesons.

A simpler way to see this is to recall the phenomenological coupling of a $W$ to two pseudoscalars, given by $W_\mu \langle[\phi,\partial_\mu \phi] T_-\rangle$ in chiral theories \cite{Gasser:1983yg, Scherer:2002tk}, with $W$, $\phi$ the fields of the $W$ and the pseudoscalar mesons and $T_-$ a matrix related to the Cabibbo-Kobayashi-Maskawa elements. The vertex $WPP^\prime$ gives rise to the operator $\epsilon_\mu (p-p^\prime)^\mu$ with $p$, $p^\prime$ the momenta of the created pair of pseudoscalars, and we see again that the $\mu=0$ component of the virtual $W$ gives rise to $p^0-p^{\prime 0}$ which carries no momentum and provides a production of the two mesons in relative s-wave, in the CM frame of these two mesons, when the masses of the particles are different, as is the case here (note, that the operator $\frac{\vec{\sigma}\cdot \vec{p}_1}{2m_1}+\frac{\vec{\sigma}\cdot \vec{p}_2}{2m_2}$ at the quark level also would vanish in the CM  when the two particles have the same mass). This mechanism is the same as the one providing the s-wave in meson baryon scattering in the local hidden gauge approach, where a vector meson is exchanged between the meson and the baryon \cite{Bansal:2015xga}, but then one has $p^0+p^{\prime 0}$ and it never vanishes.

\subsection{Coalescence production of the $D_{s0}^{*}(2317)^+$ resonance}
\begin{figure}[b]
\begin{tabular}{c}
\includegraphics[scale=0.35]{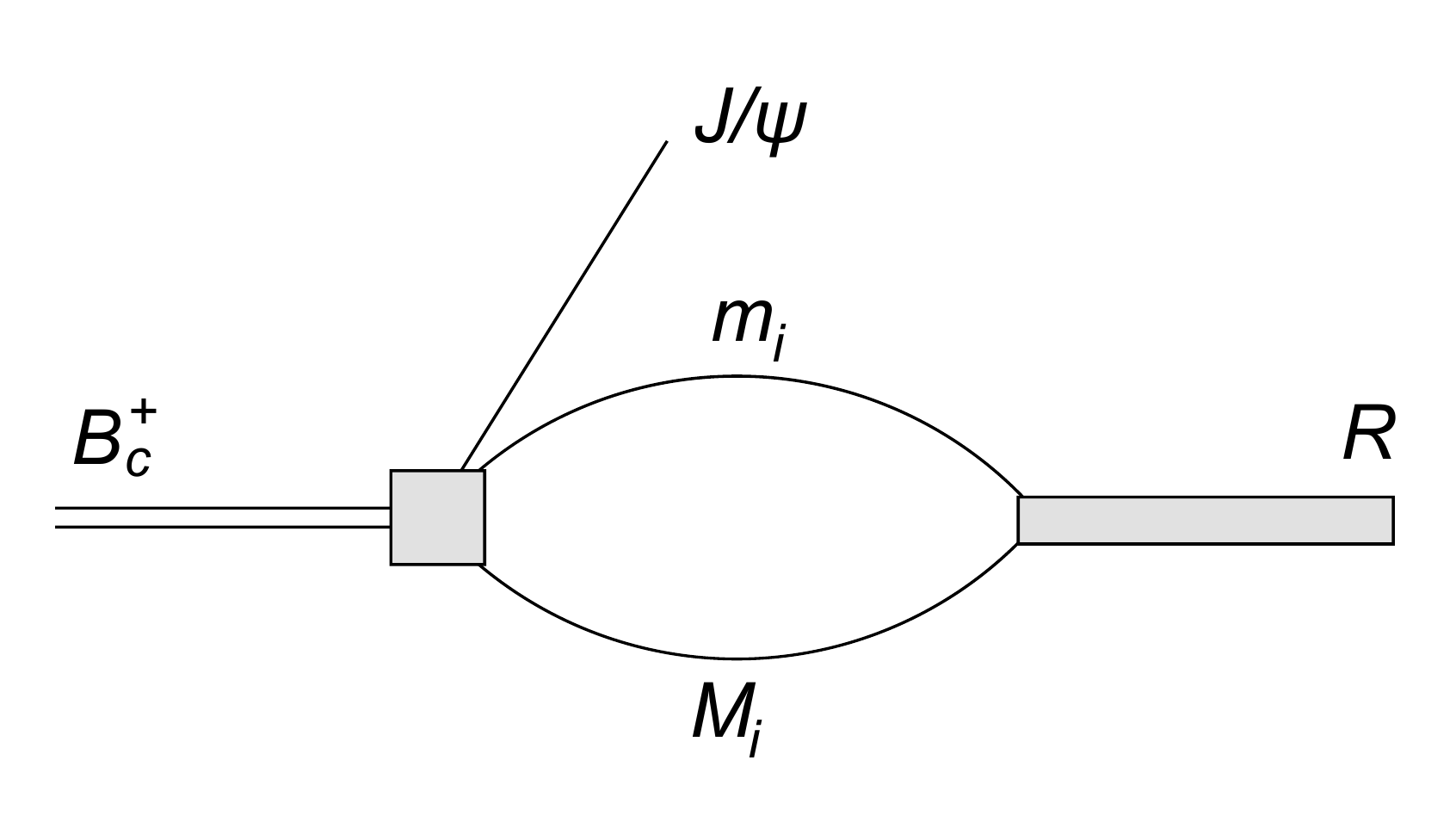}
\end{tabular}
\caption{Diagrammatic presentation of $B_c^+\to J/\psi R$. One sums over all intermediate mesons $m_i$ and $M_i$ that generate the resonance ($K^+D^0$, $K^0D^+$ and $\eta D_s^+$).\label{fig:BcJpsiDs0}}
\end{figure}

In the former subsection we have studied the production of $DK$ in the final state. Here we study the production of the resonance $D_{s0}^{*}(2317)^+$ under the assumption that it is dynamically generated from the $DK$ and $\eta D_s$ channels. Diagrammatically, the reaction proceeds as shown in Fig. \ref{fig:BcJpsiDs0}.

The amplitude for the production of the resonance $R$ (in this case the $D_{s0}^{*}(2317)^+$) is given by
\begin{eqnarray}
t(B_c^+\to J/\psi R)&=&\left.V_p\sum_ih_iG_ig_{i}\right|_{pole}\nonumber\\
&=&\left. {\sqrt{3}A} p_{J/\psi}\cos\theta \sum_ih_iG_ig_{i}\right|_{pole},\label{eq:aa}
\end{eqnarray}
where i sums over $K^+D^0$, $K^0D^+$, $\eta D_s$, and $g_{i}$ is the coupling of the resonance to the channel $i$ defined such that the scattering amplitude around the pole reads as 
\begin{eqnarray}
T_{ij}=\frac{g_ig_j}{z-z_R},
\label{Eq:coupling}
\end{eqnarray}
with $i,j=D^+K^0, D^0K^+, D_s^+\eta$. The variables $z$ and $z_R$ are the value of the complex $s$ and the resonance position, respectively.
Once again we define $\tilde{t}(B_c^+\to J/\psi R)={t}(B_c^+\to J/\psi R)/{V_p}$.
Eqs. (\ref{eq:aa}) and (\ref{eq:am_3body}) are different, but under the assumption that the resonance is dynamically generated by the channels included in Eq. (\ref{eq:am_3body}), the two expressions are related and are fixed, up to the common factor $V_p$. The width for the production of the resonance $R$, irrelevant of which decay channel it has, is given by 
\begin{eqnarray}
\Gamma(B_c^+\to J/\psi R)=\frac{A^2}{8\pi}\frac{1}{m_{B_c^+}^2}\left|\tilde{t}(B_c^+\to J/\psi D_{s0}^{*}(2317)^+)\right|^2\left. p_{J/\psi}^3\right|_{pole}.\nonumber\\\label{eq:bb}
\end{eqnarray}
It is then interesting to study the ratio \cite{Liang:2015twa}
\begin{eqnarray}
\frac{d\tilde{\Gamma}}{dM_{inv}}&=&M_R^2\frac{(d\Gamma/dM_{inv})/p_{J/\psi}^3\tilde{p}_{DK}}{\Gamma(B_c^+\to J/\psi R)/\left. p_{J/\psi}^3\right|_{pole}}\nonumber\\
&=&\frac{M_R^2}{4\pi^2}\frac{\left|\tilde{t}(B_c^+\to J/\psi D^+K^0)\right|^2}{\left|\tilde{t}(B_c^+\to J/\psi D_{s0}^{*}(2317)^+)\right|^2}\nonumber\\
&=&\frac{M_R^2}{4\pi^2}\frac{|h_{D^+K^0}+\sum h_iG_it_i|^2}{|\sum {h_i}G_ig_{i}|^2|_{pole}},\label{3b}
\end{eqnarray}
where the factor $M_R^2$ is put in the formula for convenience in order to have a dimensionless quantity. In this ratio the common factor $V_p$ (or $A$) cancels and we obtain a magnitude with no free parameters, tied to the nature of the $D_{s0}^{*}(2317)^+$ as a dynamically generated resonance.

\section{results}
\begin{figure}[t]
\begin{tabular}{c}
\includegraphics[scale=0.33]{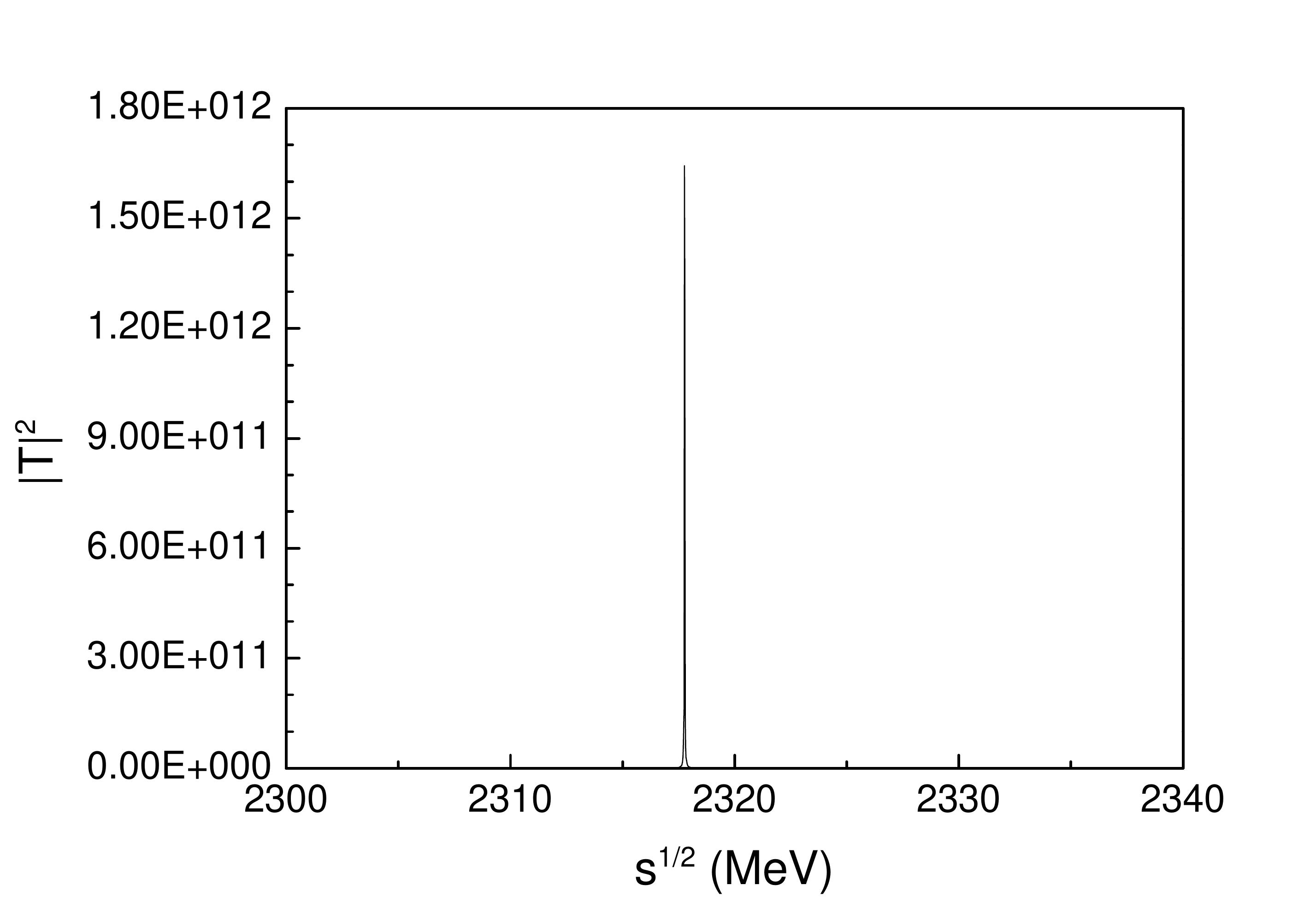}
\end{tabular}
\caption{$|T|^2$ for the $DK$ scattering depending on the center of mass energy of the $DK$ system.\label{fig:2317pole}}
\end{figure}

As it is mentioned in section II, the $G_i$ function is calculated analytically by dimensional regularization. In this paper the parameter $\alpha(\mu)$ is fixed as -1.265 and $\mu$ as 1.5 GeV, in order to get the resonance $D_{s0}^{*}(2317)^+$ from the $DK$ and $\eta D_s$ interaction. In Fig. \ref{fig:2317pole}, we show the squared amplitude of the $D^+K^0\to D^+K^0$ scattering depending on the invariant mass of $DK$ system, where the peak position appears at the mass of the $D_{s0}^{*}(2317)^+$ resonance. 
As we can see, there is no width for the state, which we obtain as a bound state. The very small width of this state comes from the decay into the isospin forbidden $\pi D_s$ channel, which we do not consider since it has a negligible role in the generation of the $D_{s0}^{*}(2317)^+$ mass.
We also calculate the couplings of the $D_{s0}^{*}(2317)^+$ to the $DK$ and the $D_s\eta$ channels, which can be extracted from the behavior of the $T$ matrix near the pole (see Eq. \eqref{Eq:coupling}. The values of the couplings that we get are $g_{D^+K^0}=g_{D^0K^+}=7.4$ GeV, $g_{D_s^+\eta}=-6.0$ GeV.
This corresponds to a coupling $-\sqrt{2}\times 7.4=-10.46$ (GeV) to the $DK$ $I=0$ channel $-\frac{1}{\sqrt{2}}(K^+D^0+K^0D^+)$. The value of $-10.46$ GeV and $-6.0$ GeV for the couplings of the $D_{s0}^{*}(2317)^+$ to $DK$ $(I=0)$ and $\eta D_s$ agree with those in \citep{Gamermann:2006nm} up to a global sign.
Note that here we take $g_{D^+K^0}$ positive.
We see then that the $D_{s0}^{*}(2317)^+$ resonance couples to the $DK$ channel more strongly.

\begin{figure}[htb]
\begin{tabular}{c}
\includegraphics[scale=0.33]{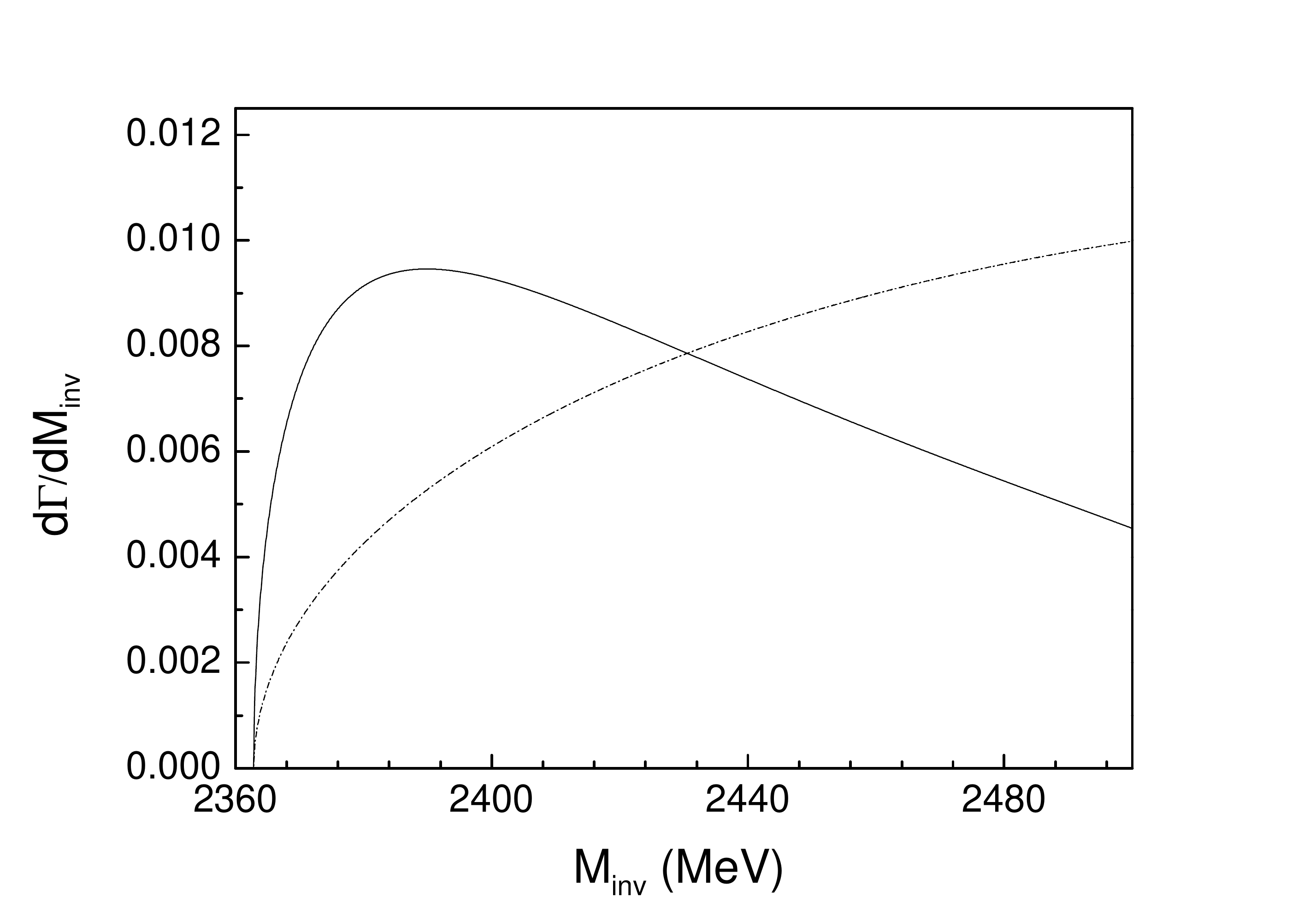}
\end{tabular}
\caption{Differential decay width for the reaction $B_c^+\to J/\psi D^+K^0$. The solid curve corresponds to $(\alpha(\mu), \mu)=(-1.265, 1.50$ GeV$)$. The dash dot curve is the phase space. Note that all the curves have been normalized in the range of $M_{inv}$ shown in the figure.\label{fig:mass distribution}}
\end{figure}

Using the values of $\alpha(\mu)$ and $\mu$ mentioned above, we can get the differential decay width for the reaction of $B^+_c\to J/\psi D^+K^0$ (see Fig. \ref{fig:mass distribution}). There the line shape of the differential decay width and the phase space have been normalized to unity over the range of the $DK$ invariant mass in the figure, which have been done in the same way as that in Ref. \cite{Albaladejo:2015kea}. The line shapes are similar to those in Fig. 4 of Ref. \cite{Albaladejo:2015kea}.

\begin{figure}[htb]
\begin{tabular}{c}
\includegraphics[scale=0.33]{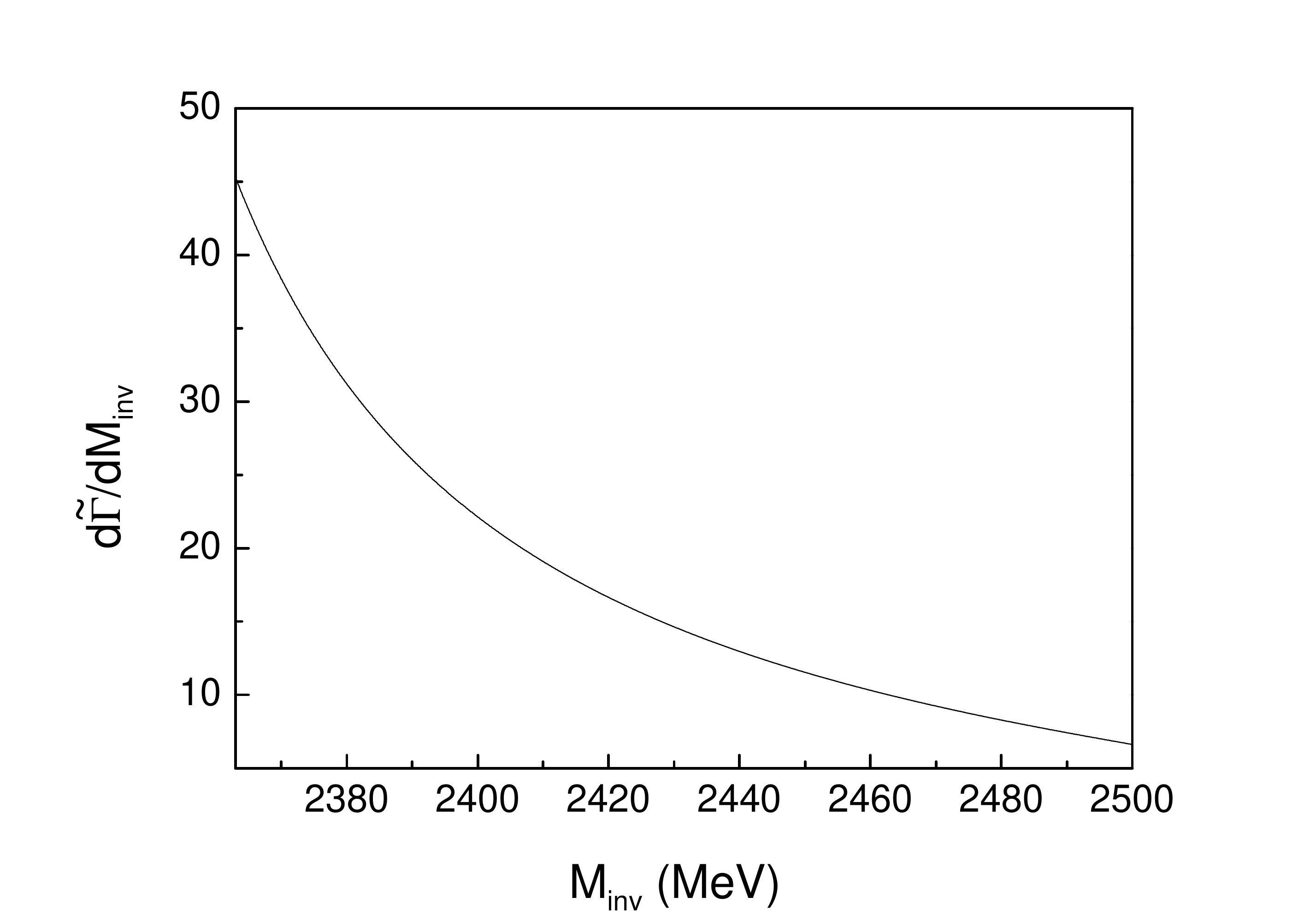}
\end{tabular}
\caption{The plot of $\frac{d\tilde{\Gamma}}{dM_{inv}}$ defined in Eq. \eqref{3b}.\label{fig:ratio}}
\end{figure}

In Fig. \ref{fig:ratio} we plot $\frac{d\tilde{\Gamma}}{dM_{inv}}$ of Eq. (\ref{3b}). We see a fall down of the distribution as a function of the $K^+D^0$ invariant mass. This is a clear indication of the presence of a resonance below threshold since we have divided the original invariant mass distribution by the phase space. Hence, essentially we are plotting $|t(B_c^+\to J/\psi D^+K^0)|^2$, which peaks at the mass of the $D_{s0}^{*}(2317)^+$ and we are seeing the tail of the resonance.

\subsection{Consideration of possible $q\bar{q}$ components}

At this point we would like to consider uncertainties tied to the possibility that the state $D_{s0}^*(2317)$ contains some $q\bar{q}$ component in addition to the $KD$ and $\eta D_s$ molecular channels. This possibility was investigated in Ref. \cite{Torres:2014vna} when analyzing the lattice QCD spectra for $KD$ and related channels, including $q\bar{q}$ components. The analysis to account for a possible $q\bar{q}$ component was done by adding a Castillejo-Dalitz-Dyson pole \cite{Castillejo:1955ed} to the potential. Hence, in the charge basis that we consider, we add to the potential of Eq. \eqref{eq:potential} an extra component
\begin{eqnarray}
\delta V_{11}=\delta V_{12}=\delta V_{21}=\delta V_{22}=\delta V=\frac{\gamma^2}{M_{inv}^2-M_{CDD}^2},\label{eq:extrapotential}
\end{eqnarray}
where the equality of these terms guarantees that we have $I=0$ and we have only considered the extra term in the important $KD$ channels, as in \cite{Torres:2014vna}. In Eq. \eqref{eq:extrapotential} $\gamma$ is a constant and $M_{CDD}$ would be associated to a bare mass of some excited $q\bar{q}$ component. With the potential $V+\delta V$ we reevaluate amplitudes and couplings. The analysis done in \cite{Torres:2014vna} gave as an output that the possible values of $M_{CDD}$ were larger than $M_K+M_D+300$ MeV. We choose $M_{CDD}=M_K+M_D+300$ MeV, but we have checked that the results do not change if we have a larger $M_{CDD}$ mass after the fitting is done. The procedure to determine the $\gamma$ parameter is the following.

First we quote from Ref. \cite{Torres:2014vna} that the probability to have the $KD$ component was, summing errors in quadrature, $P(KD)=(72\pm 14)\%$. Hence we take three cases, corresponding to $P(KD)=58\%$, $72\%$, $86\%$. For each of these cases we make a fit of the subtraction constant $\alpha(\mu)$ of Eq. \eqref{eq:loopfunction} and $\gamma$ to obtain the binding at 2317.8 MeV and $P(KD)$ equal to any of the former fractions. The $P(KD)$ probability is given by \cite{Gamermann:2009uq, Hyodo:2008xr, Hyodo:2013nka}
\begin{eqnarray}
P(KD)=\left.-\sum_{i=1,2}g_i^2\frac{\partial G_i}{\partial s} \right|_{pole}
\end{eqnarray} 
with $i=K^+D^0$, $K^0D^+$.

\begin{figure*}[htb]
\includegraphics[scale=0.7]{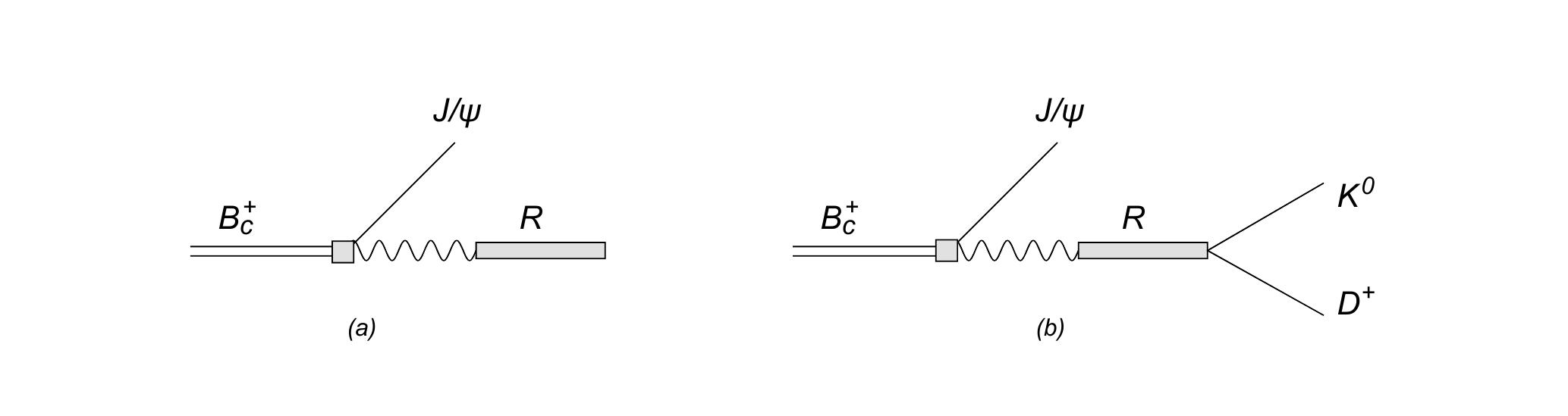}
\caption{Diagrams for coalescence (a) and $K^0D^+$ production (b) through coupling of the $B_c^+J/\psi$ vertex to a possible $q\bar{q}$ component.\label{fig:coalescence}}
\end{figure*}
\begin{figure*}[tb]
\includegraphics[scale=0.2]{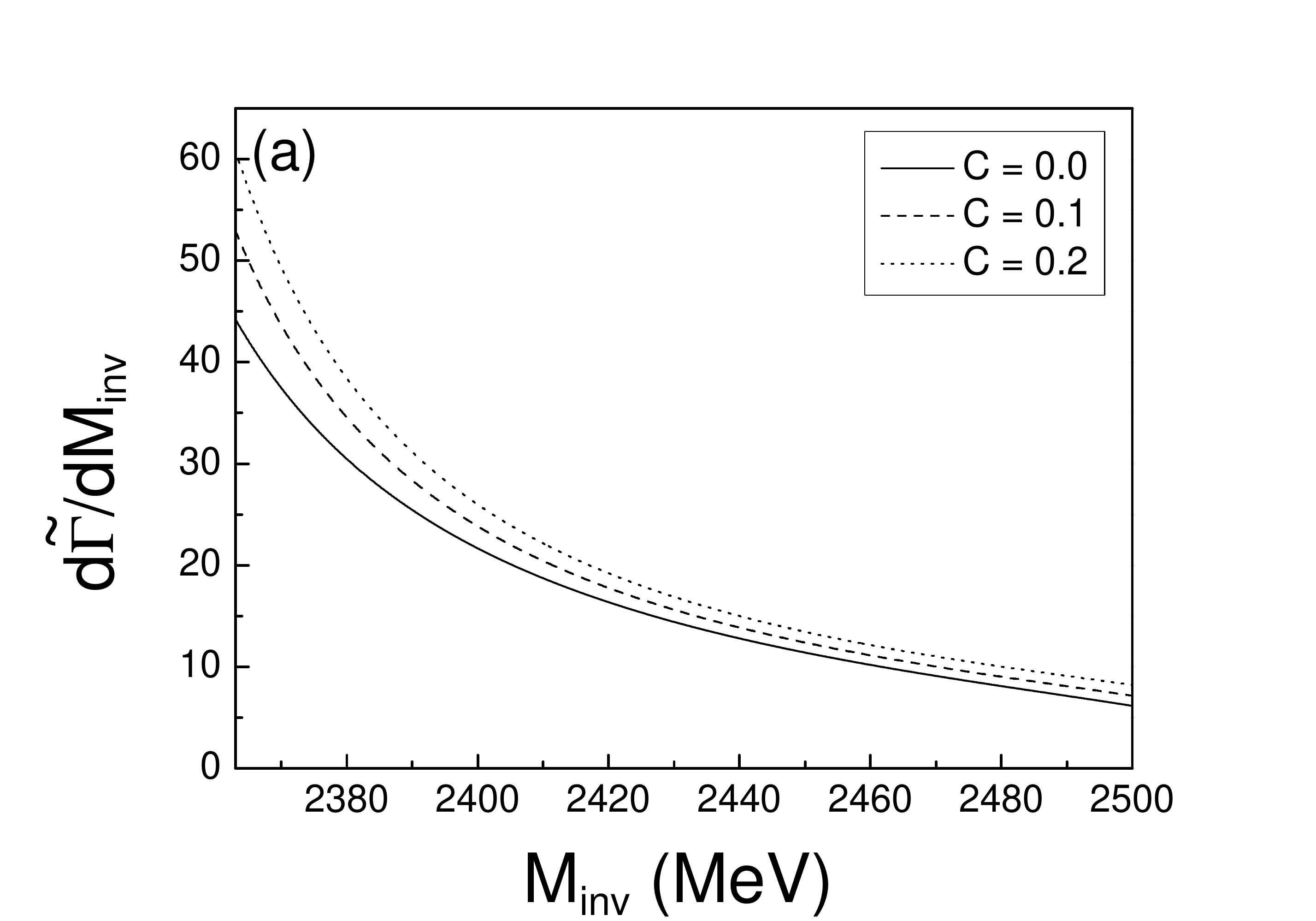}
\includegraphics[scale=0.2]{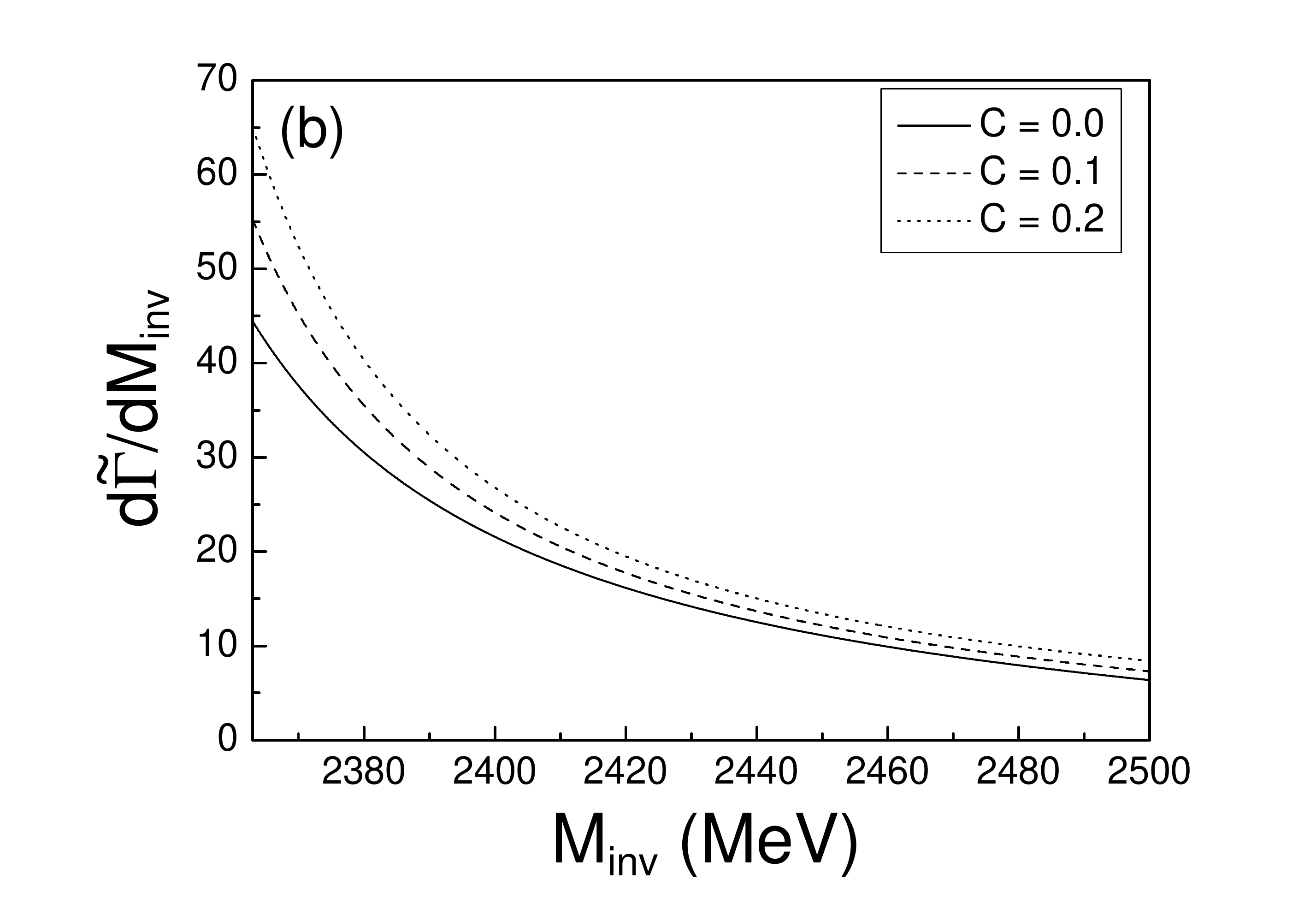}
\includegraphics[scale=0.2]{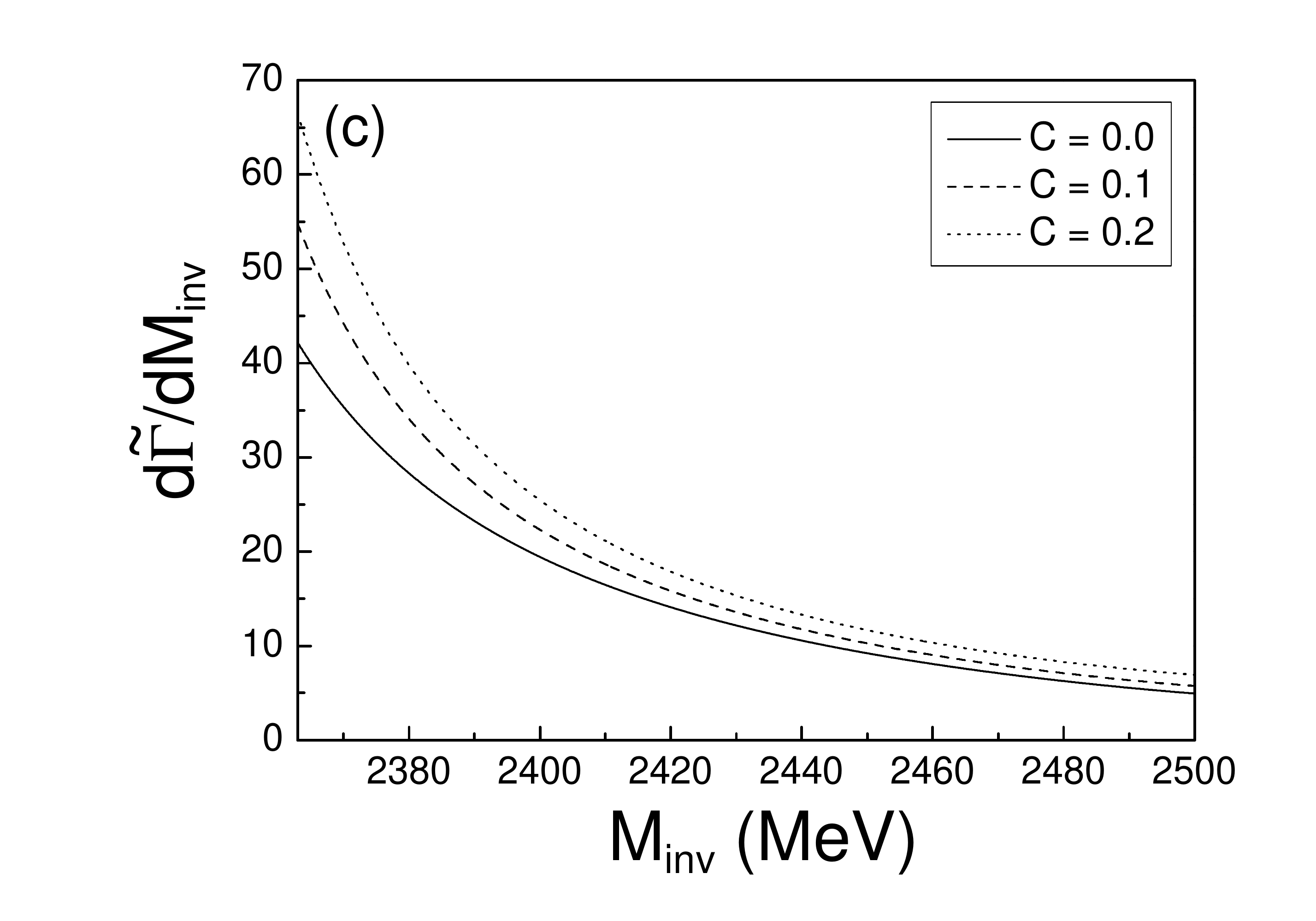}
\caption{The plot of $\frac{d\tilde{\Gamma}}{dM_{inv}}$ considering the contribution of Fig. \ref{fig:coalescence}. (a), (b) and (c) are corresponding to $P(KD)=0.58$, $0.72$ and $0.86$ respectively. \label{fig:new_ratio}}
\end{figure*}

Next we take into account that the new component not only modifies couplings and amplitudes of $KD$ but in the decay that we consider there can be a direct coupling to this component. This is depicted in Fig. \ref{fig:coalescence} (a) for the coalescence process and in Fig. \ref{fig:coalescence} (b) for the $K^0D^+$ production. We must add this contribution to the former one, but we have here two extra couplings which are unknown, the $B_c^+J/\psi$ coupling to the $q\bar{q}$ component and the resonance coupling to this component. Let us write the product of the two in terms of an unknown constant $C$, as $V_pM_R^3C$ for dimensional reasons. Then the amplitudes equivalent to Eq. \eqref{eq:am_3body} and \eqref{eq:aa} are now: 
\begin{eqnarray}
t(B_c^+\to J/\psi R)&=&\left. V_p\left(\sum_i h_iG_ig_i+C\frac{M_R^3}{M_R^2-M_{CDD}^2}\right)\right|_{pole}\label{eq:3aaa}\\
t(B_c^+\to J/\psi D^+K^0)&=& V_p\left(1+\sum_i h_iG_ig_it_{i1}+C\frac{M_R^3}{M_R^2-M_{CDD}^2}\right.\nonumber\\
&&\times\left. \frac{g_{D^+K^0}}{M_{inv}^2-M_R^2}\right)
\end{eqnarray}
where $M_R$ is the mass of the $D_{s0}^+(2317)$.

The parameter $C$ is unknown, but with $KD$ and $\eta D_s$ nearly exhausting the sum rule $-\sum_{i=1}^3g_i^2\frac{\partial G_i}{\partial s}\simeq 1$, we shall choose it in such a way that the $C$-term in Eq. \eqref{eq:3aaa} gives a weight of $0$, $0.1$ or $0.2$ of the $\sum_i h_iG_ig_i$ term. With this the relative strength of this term on $|t|^2$, and hence in $\Gamma$, would be $0$, $21\%$ and $44\%$ respectively. We think this is a wide margin given the small room left for such component in the sum rule. We should expect this contribution to be of the order of the uncertainty of $14\%$ in $P(KD)$ that we have quoted before, but we take a wider margin.

When this is done we get the results that we show in Fig. \ref{fig:new_ratio}. Compared to the former results in Fig. \ref{fig:ratio}, we see small changes. For weight $C=0$ in Fig. \ref{fig:new_ratio} (b) where $P=0.72$ we get a reduction of $\frac{d\tilde{\Gamma}}{dM_{inv}}$ about $10\%$ with respect to the former results. For the case that the relative weight of the new component is $0.1$ in the amplitude we get results about $5\%$ bigger than before. And for relative weight 0.2 we get an increase of about $20\%$. We also can see in Figs. \ref{fig:new_ratio} (a) and (c) for weights $P(KD)=0.58$ and $P(KD)=0.86$ that the results change only in about $5\%$ with respect to those for $P(KD)=0.72$. This is a margin of uncertainty that we can assume, but the main features discussed above remain.

\section{conclusion}
In this paper we have studied the $B^+_c$ decay into $J/\psi D^+K^0$. The mechanism is: $B^+_c$ decays into $J/\psi$ and the quark pair $c\bar{s}$ via weak interaction; then the quark pair $c\bar{s}$ hadronizes into $D^+K^0$, $D^0K^+$ or $D_s^+\eta$ components which can interact among themselves generating the $D_{s0}^{*}(2317)^+$ resonance. In the scheme of the chiral unitary approach, we are able to choose the proper parameters $\alpha(\mu)$ and $\mu$ appearing in the loop function by matching the pole position of the $D_{s0}^{*}(2317)^+$. If $\alpha(\mu)=-1.265$ and $\mu=1.5$ GeV, the couplings of $DK$ and $D_s\eta$ channels are $g_{D^+K^0}=g_{D^0K^+}=7.4$ GeV and $g_{D_s^+\eta}=-6.0$ GeV, respectively. 
Later we have calculated the differential decay width of the reaction $B^+_c\to J/\psi D^+K^0$. 
One can appreciate that the shape of the distribution peaks closer to the $DK$ threshold than the phase space, indicating the coupling of $DK$ to a resonance below threshold (the $D_{s0}^{*}(2317)^+$ in this case). We also evaluated the rate of production of the $D_{s0}^{*}(2317)^+$ resonance and then constructed the ratio of $d\Gamma/dM_{inv}(B_c^+\to J/\psi D^+K^0)$ to the width for $D_{s0}^{*}(2317)^+$ production, where the unknown factor $V_p$ of our theory cancels. The new normalized distribution obtained is then a prediction of the theory, only tied to the fact that the $D_{s0}^{*}(2317)^+$ is dynamically generated from the $DK$ and $\eta D_s$ channels. We also evaluated the possible contribution of genuine $q\bar{q}$ components taking information from the lattice QCD results and found it to be small. As to the feasibility of the reaction we think this is at reach in present facilities. Indeed in the PDG \cite{Agashe:2014kda} one finds half of the known decay channels of the $B_c^+$ going to a $J/\psi$, one has also decays into $J/\psi$ and three pions, $J/\psi$ plus two kaons and one pion, and also decays into $J/\psi D_s^+$ and $J/\psi D_s^{*+}$. The study done here, showing how one can learn about the nature of the $D_{s0}^{*}(2317)^+$ from the measurements proposed, should serve as an incentive to perform these experiments in the near future.

\section*{Acknowledgments}

This work is partly supported by the Spanish Ministerio de Economia y Competitividad and
European FEDER funds under the contract number FIS2011-28853-C02-01 and FIS2011-28853-
C02-02, and the Generalitat Valenciana in the program Prometeo II-2014/068. We acknowledge
the support of the European Community-Research Infrastructure Integrating Activity Study
of
Strongly Interacting Matter (acronym HadronPhysics3, Grant Agreement n. 283286) under the
Seventh Framework Programme of EU.

\bibliographystyle{plain}

\end{document}